\documentstyle[aps,preprint,tighten,floats,epsf,rotate]{revtex}

\begin{document}
\draft

\title{Iso-curvature fluctuations through axion trapping by cosmic string wakes}
\author{Biswanath Layek \footnote{e-mail: layek@iopb.res.in}}
\address{Institute of Physics, Sachivalaya Marg, Bhubaneswar 751005, 
India}
\maketitle
\widetext
\parshape=1 0.75in 5.5in
\begin{abstract}

 We consider wake-like density fluctuations produced by cosmic strings at
the quark-hadron transition in the early universe. We show that low momentum
axions which are produced through the radiation from the axionic string 
at an earlier stage, may get trapped inside these wakes due to delayed 
hadronization in these overdense regions.  As the interfaces, bordering 
the wakes, collapse, the axions pick-up momentum from the walls and finally 
leave the wake regions. These axions thus can produce large scale 
iso-curvature fluctuations. We have calculated the detailed profile of 
these axionic density fluctuations and discuss its astrophysical consequences.

\end{abstract}
\vskip 0.125 in
\parshape=1 0.75in 5.5in
\pacs{PACS numbers: 14.80.Mz, 98.80.Cq, 12.38.Mh}
Key words: {cosmic string, axion, density fluctuation, quark-hadron 
transition} 
\narrowtext

\section{Introduction}

If QCD phase phase transition is of first order in nature then there are
many important consequences in the context of early universe as well as
in heavy ion collisions.  In the conventional picture of a first order phase 
transition, the transition happens through homogeneous nucleation 
of hadronic bubbles in the background of QGP phases. Hadronic bubbles
thus formed will keep expanding, coalesce and 
eventually complete the phase transition.  One of the important consequences 
in this scenario, in the context of the early universe, 
is the concentration of baryon numbers inside the shrinking 
QGP bubbles which eventually can even form quark nuggets \cite{wtn} due to 
high concentrations of baryon number inside the shrinking QGP phases. It was 
also shown that concentration of baryon number in the QGP phase can produce 
large baryon fluctuations which can survive against various dissipative 
processes \cite{fuller} upto nucleosynthesis epoch and can alter the primordial
abundances \cite{ibbn1,ibbn2}.  Ignatius and Schwarz studied \cite{inhm} 
the presence of density fluctuations (those arising  from inflation) at  
quark-hadron transition and showed that it will lead to splitting of 
the region in hot and cold regions with cold regions converting to 
hadronic phase first. Baryons will then get trapped in the (initially) 
hotter regions. Estimates of sizes and separations of such density 
fluctuations were made in ref.\cite{inhm} using COBE measurements of 
the temperature fluctuations in CMBR. In our earlier 
work \cite{layek1,layek2}, we considered the effect of cosmic string 
induced density fluctuations on quark-hadron transition and showed that 
such density fluctuations at QCD epoch can lead to formation of extended 
planar regions of baryon inhomogeneity. Detailed profile of baryon 
inhomogeneities and it's effect on standard big bang nucleosynthesis 
were discussed in ref.\cite{layek2}.

  Here, we study another aspect of the QCD phase transition in the context
of early universe if the transition is of 1st order in nature, that is, axion 
inhomogeneity generation during quark-hadron transition in the presence of
cosmic string induced density fluctuations. Axion was  first 
introduced to solve strong CP problem \cite{pq}  and has been extensively 
investigated primarily due to its possible role as a dark matter candidate. 
Recently it has been discussed \cite{hind} that due to the larger
mass of axion in the hadronic phase compared to the QGP phase, axions will
get trapped inside the shrinking QGP bubbles at the quark-hadron transition
(somewhat in the similar manner as baryon trapping occurs in shrinking
QGP bubbles \cite{wtn}). The axions which are initially trapped inside 
the QGP phase will gradually pick up momentum from the walls of the 
shrinking bubbles and leave the QGP phase \cite{hind}. These axions which 
are left behind as the QGP bubbles collapse can form iso-curvature 
fluctuations. The minimum momentum $p_{min}$ required to leave the QGP 
regions is given by,

\begin{equation}
p_{min}= \sqrt {(m_{ah}^2 - m_{aq}^2)}\equiv \Delta m 
\end{equation}

Where, $m_{ah}$ and $m_{aq}$ are the axion masses in the hadron phase and QGP
phase respectively. 

 We study this phenomenon of axion trapping when cosmic strings are present
in the early universe. Cosmic strings are generically produced by GUT
phase transitions. Further, it has been recently shown that superstring
theories also lead to cosmic strings with acceptable parameters 
\cite{sstrng}(see also ref. \cite{kibble}). It therefore becomes important 
to investigate how such cosmic strings can affect the universe. (It is clear
by now from latest WMAP data that contributions of cosmic strings and other 
topological defects to structure formation is insignificant \cite{struct}).
We have shown earlier that the process of hadronization will be delayed in 
the  cosmic string wakes at the quark-hadron transition \cite{layek1,layek2}.   
The mechanism of shrinking quark phases is  very different in our 
model compared to the conventional scenario based on homogeneous bubble 
nucleation. Geometry of the collapsing interface in our case is of sheet like
planar structure unlike spherical in the conventional case.  Thus, the 
iso-curvature fluctuations produced, as the axions leave the wake region, 
will spread in a sheet like regions in contrast to the spherical clumps as
studied by Hindmarsh \cite{hind}.  It will be interesting to study the 
subsequent evolution of large fluctuations with such geometrical 
structure and its astrophysical consequences.
The paper is organized in the following manner. In sec. II, we briefly 
review the axion production from cosmic axionic string and it's temperature
dependent mass following ref. \cite{hind}. Sec. III discusses the nature of
density fluctuations as expected from straight cosmic string moving through
relativistic fluid. Density fluctuations produced by the moving
"wiggly" string is also discussed in this section.  In Sec. IV, the dynamics
of quark-hadron transition in the presence of string wakes 
and the mechanism of axion trapping in our model have been discussed.  In Sec.
V, we give our result. Summary and possible astrophysical consequences have
been discussed in the final section VI.

\section{Axion Production from Cosmic Strings}

The presence of the so called $\theta$ term in the QCD Lagrangian
violates CP invariance. The most stringent bound on this $\theta (< 10^{-9})$
comes from the electric dipole moment measurement of neutron.
Such extremely small value of $\theta$ in the strong interaction theory 
is called "strong CP problem". One of the most elegant solutions
of this problem was first proposed  by Peccei and Quinn \cite{pq} by 
introducing an axial $U(1)_{PQ}$ symmetry. The spontaneous breaking of this 
symmetry at some scale $\eta_a$ can cause $\theta$ to settle down at some 
vacuum value. At the QCD scale instanton effects lead to explicit breaking
of this $U(1)_{PQ}$ symmetry forcing $\theta$ to settle down at the true
vacuum $\theta = 0$. However, due to spontaneous breaking of $U(1)_{PQ}$
symmetry, the phase of the field can wind around the vacuum manifold 
non-trivially in the physical space, forming the so called axionic string.
After formation the strings are stuck in the plasma and are stretched by
the Hubble expansion. During this time the dominant mechanism of dissipating
energy is via heat due to the large frictional force exerted by the background
plasma. However, with time the plasma becomes dilute and below a temperature 
$T_*$ which is given by \cite{sikivie},

\begin{equation}
T_*=2 \times {10^7} GeV ({\eta_a \over 10^{12} GeV})^2
\end{equation}

the strings move freely. The corresponding time $t_*$, can be obtained from
the following time-temperature relation,

\begin{equation}
T^{2}= {0.3 \over \sqrt{g_*}} {m_{pl} \over t}
\end{equation}

and is given by,

\begin{equation}
t_*= {0.3 \over \sqrt{g_*}} {m_{pl} \over T_*^{2}} \simeq 1.8 
\times 10^{-21} sec ({{10^{12} GeV} \over \eta_a})^4
\end{equation}

Here $g_* = 106.75~$ is the number of degrees of freedom relevant
at temperature $T_*$ (taking all the species in the standard model).  After 
$t_*$, the string will lose energy through radiation of (pseudo) Goldstone 
bosons which are called axions.  The axion is not truly a Goldstone Bosons. 
It will acquire mass due to the axial anomaly \cite{thooft} once the instanton
effect turns on. In the high temperature phase ($T> T_c$), the mass of 
axions arises from the $\theta$-dependent part of the free energy density which
have been calculated by Preskil et al.  \cite{mass} in the dilute-instanton-
gas approximations. For three quark flavors appropriate for T not too high compared to $T_c$, they have determined the mass as,

\begin{equation}
m_{aq}(T) \simeq 2 \times 10^{-2} [{{\Lambda ^{2} \over \eta_a}}] 
[{m_u m_d m_s \over {\Lambda ^{3}}} ^{1/2}][{\Lambda \over {\pi T}}]^{4} 
[9 ln {\Lambda \over {\pi T}}]^{4}
\end{equation}

Where, $m_u, m_d, m_s$ are the current quark masses of u, d and s quark 
respectively and $\Lambda \equiv \Lambda_{QCD} \sim 200 MeV$ is the QCD scale.  
Taking critical temperature $T_c \sim 150 MeV$ and putting the values of
u, d and s quark masses (see ref.\cite{hind} for details), we get the axion
mass at the onset of QCD transition and can be written in terms of axionic
string formation scale as,

\begin{equation}
m_{aq}(T_c)= 5.3 \times 10^{-7} eV ({{10^{12} GeV} \over \eta_a }) 
\end{equation}

At low temperature in the hadron phase, the axion mass is calculated 
from the current-algebra technique which gives the value \cite{mass},

\begin{equation}
m_{ah}= {{(m_u m_d)}^{1/2} \over m_u + m_d} {m_{\pi} f_{\pi} \over \eta_a}
= 6.04 \times 10^{-6} eV ({{10^{12} GeV} \over \eta_a })
\end{equation}

Here, $m_{\pi}$ and $f_{\pi}$ are the mass and decay constant of pions 
respectively. The axion mass being temperature dependent at high temperature
phase as in Eq.(5), it's very light and ultra-relativistic at very high 
temperature. It becomes dynamically significant at the temperature at which
compton wave length of the axions fall within the horizon. This is the time 
($\tilde t$), at which axion string will not be able to oscillate sufficiently
to radiate axions due to damping of domain walls attached to each string. 
From the time temperature relation Eq.(3) one can get the time $\tilde {t}$,
from the following relation (neglecting the logarithmic dependence on T in 
Eq.(5)), 

\begin{equation}
m_{aq}(\tilde {t})=m_{aq}(t_c) ({\tilde t \over t_c})^2= d_h(\tilde t) = 
(2 \tilde {t})^{-1}
\end{equation}

Where, $m_{aq}(t_c)$ is the mass of the axion at the onset of QCD phase 
transition obtained from Eq.(5). Thus, one gets the time $\tilde {t}$ as,

\begin{equation}
\tilde {t}= {1 \over 2}[m_{aq}(t_c) t_c]^{-1/3} t_c \simeq 1.6 
\times 10^{-3} sec ({\eta_a \over {10^{12} GeV}})^{1/3} 
({t_c \over sec})^{2/3}
\end{equation} 

Here,  we will study the fluctuations in the density of axions which are 
produced through the radiation from the axionic string during the era, 
$t_*<t<\tilde t$ as in ref. \cite{hind}. During this time one can assume 
that string network will enter into the scaling regime which has been 
discussed extensively in literature (see, ref. \cite{str}).  Under the 
scaling assumption the spectrum of axions in the comoving momentum 
range $k^*$ and $\tilde k$  can be considered to be flat \cite{hind}. We 
will take the number density distribution of axions resulting from axionic 
string decay in the comoving momentum range $k^*(={(R(t_*) \over t_*})$ and 
$\tilde {k} (={R(\tilde t) \over \tilde t})$ as given in ref. \cite{hind} as,

\begin{equation}
n_k dk = R(\tilde {t}) G \eta_a^{2} ln(\eta_a \tilde t) {\rho _c} dk/k^{2}
\end{equation}

Where, $R(\tilde t)$ is the scale factor at time $\tilde t$, and $\rho_c$ is
the corresponding critical density of the universe. We will consider the above
density distribution with suitable dilution factor due to expansion of the 
universe inside the wake-like overdensity density regions produced at QCD
epoch as discussed in the following section.

\section{Density Fluctuations arising due to Long Cosmic Strings}

In this section we will review the formation of density fluctuations due
to  cosmic string moving through relativistic fluid. First, we will
discuss the case for a long straight string assuming it does not have any
small scale structure. Though, simulations \cite{wiggle1} on the string 
network in the expanding universe show that the long cosmic string possesses 
substantial amount of small scale "wiggly" structures running along the 
strings. The presence of such wiggles makes the string heavier and  
slows down the motion of the string as a whole. However, as
we discuss below, even the presence of such wiggles along the string will not
alter the order of magnitude estimate of density fluctuations produced 
at QCD epoch. Therefore, for simplicity of presentation we will consider 
the case of straight (i.e., with no small scale structure)  strings
first and then we briefly discuss for the case of wiggly strings.

To discuss the generation of density fluctuations by moving straight 
string, we take the metric describing the space-time around the string (lying
along z-axis) as \cite{mtrc},

\begin{equation}
ds^2 = dt^2 - dz^2 - dr^2 -  (1 - 4G\mu)^2 r^2 d\psi^2 .
\end{equation}

Here, $\mu$ is the mass per unit length (equal to string tension) of the 
straight string. 
 This metric
can be put in Minskowski form by defining angle $\psi^\prime = (1-4G\mu)\psi$, 
where new angle  $\psi^\prime$ varies between 0 and $(1-4G\mu)2\pi$ instead of
0 and $2\pi$. A test particle at rest with respect to the string does not feel
any gravitational force. However, when the string moves with velocity $v_{st}$,
the nearby matter gets an impulsive velocity along the direction of the surface
swept by the moving string. For collisionless particles, the resulting 
impulsive velocity \cite{str} is $v_{impul} = 4\pi G\mu v_{st} \gamma_{st}$. 
This effect is responsible for the formation of wake \cite{wake} and for
collisionless particles, this leads to planar structure of density
fluctuation. In this case, the density fluctuations  will be of order unity 
with opening angle of the wake being equal to the deficit angle $8\pi G\mu$. 
However, at QCD epoch the universe was dominated by the relativistic plasma.
So, to study the density structure of the wake at that epoch, the  description
of matter should be taken in terms of a relativistic fluid rather than 
collisionless particles. Generation of density fluctuations due to a cosmic
string moving through  a relativistic fluid has been analyzed in the literature 
\cite{shk1,shk2,shk3}, where it has been shown that for supersonic velocities of
string, a wake-like shock is formed behind the string. We will follow ref.
\cite{shk3}, where the equations of motion of a relativistic fluid
are solved in the string space-time (Eq.(11)), and both subsonic
and supersonic flows are analyzed. One finds that for
supersonic flow, a shock develops behind the string, just as
in the study of ref. \cite{shk1,shk2}. Following ref. \cite{shk3}, the
resulting density fluctuations can be expressed  in terms of four-velocity 
and given by,

\begin{equation}
{\delta\rho \over \rho} \simeq {16\pi G\mu u_f^2(1+u_s^2) \over
3u_s\sqrt{u_f^2 - u_s^2}}, \qquad sin\theta_w \simeq {u_s \over u_f} ,
\end{equation}

\noindent where $u_f (= v_f/\sqrt{1-v_f^2})$ and $u_s (= v_s/\sqrt{1-v_s^2})$
are the four velocity of fluid (in the string rest frame) and four velocity
of sound respectively, with $v_s$ being the sound speed. $\theta_w$ is the
angle of the shock.  In this case, when string velocity $v_f$ is 
ultra-relativistic, then one can get strong over densities (of order 1)
and the angle of the wake approaches the deficit angle $\simeq 8\pi G\mu$.
To discuss the effect of density fluctuations on QCD phase transition we will
follow our previous work \cite{layek1,layek2} and take the sample value
of density fluctuations and the angle of the shock as,

\begin{equation}
\theta_w \simeq 20^0, \qquad  \delta \rho/\rho \simeq 3 \times 10^{-5} .
\end{equation}

These values are obtained from Eq.(12) for string velocity of 0.9, 
sound velocity $v_s = 1/\sqrt{3}$ and $G\mu$ $\sim 10^{-6}$ (for 10$^{16}$
GeV GUT strings).

  Here, we should mention that the axion being very weakly interactive \cite{hind}, they can be considered to be a fluid 
consisting of collisionless particles. Therefore, description of wake formation
for collisionless particles as mentioned above can be operative for axions also
and can lead to formation of 
axionic density fluctuations $({\delta \rho \over \rho})_{axion}$ of order unity 
with angle of the wake, $\theta_w ~ \sim 8\pi G\mu$. These over dense axions
could be concentrated within very thin sheet like region of thickness  
$8\pi G\mu ~ d_H \sim ~ 1 cm$ ($d_H \sim ~ 10 $ km being the horizon 
size at $T_c = 150$ MeV). (One has to properly account for the wavelength of
axions.) However, we will see below that the magnitude of the density fluctuations 
produced through axion trapping
(which is our main topic of discussion in this paper) by the collapsing 
interfaces will  be very large compared to the above density fluctuation.
Essentially, this thin sheet like region will be trapped initially inside 
the wake of larger thickness which is produced by the formation of shock by 
moving string through relativistic fluid (as discussed above). Ultimately, 
axions will also escape from this thin wake and contribute in producing overall iso-curvature 
fluctuations. However, their effect will be very insignificant. Even the presence 
of wiggles along the string as we discuss below, will not change the order of 
magnitude estimate of such axionic density fluctuations. Therefore, to make the
physical picture of axion trapping simpler, we will not consider these kind of 
fluctuations in this work and we will focus only on iso-curvature fluctuations 
produced through trapping of axions inside the collapsing interfaces of the 
wake.

  So far we have discussed density fluctuations produced by the straight
string assuming it does not have any small scale structure. But, as we 
mentioned 
above, the string has small-scale wiggly structures \cite{wiggle1} running 
along the string. The presence of such wiggles increase the average energy
density of the string and makes the strings to move slower. For physics at 
large scales (compared to the scale of wiggles  $\l_w$, which is expected to be
set by the gravitational-radiation from the string \cite{wiggle2} and given as, 
$l_w \simeq \Gamma G\mu t$ where $\Gamma \sim$ 100), the effect of wiggles is taken
into account by defining the effective mass per unit length $\mu_{w}$ and the string 
tension $T_w$. Unlike the case of straight string, these 
quantities for perturbed string are not equal and they are related by 
the equation of state \cite{wiggle2}, $ \mu_w T_w = \mu^2$. Due to the 
presence of these small scale structure, the static wiggly strings
develop Newtonian gravitational field and a test particle at rest with respect
to such perturbed string experiences attractive gravitational force.
The metric around the wiggly string and density structure of the wake for 
collisionless particles have been studied and discussed in the literature
\cite{str,wiggle2}. It turns out that, for moving wiggly string, total 
impulsive velocity imparted on the particles towards the surface swept out 
by the string can be written as \cite{wiggle2},

\begin{equation}
v_{impul} =  4\pi G\mu_w v_{st} \gamma_{st} + {{{2\pi G (\mu_w - T_w)}} \over 
{v_{st} \gamma_{st}}}.
\end{equation}

Where, first term is the usual velocity impulse arises due to the conical 
structure of the space-time around the wiggly string, with $\mu$ being replaced by the 
effective mass per unit length $\mu_w$. The second term arises due to the 
Newtonian gravitational field developed by the wiggliness of the string. Using 
the value of $\mu_w \sim 1.9 \mu$ and $T_w \sim 0.5 \mu$ (see ref. \cite{str}), we
see from Eq.(14) that if the string moves with ultrarelativistic speed, then the 
first term will dominate over the gravitational term. The rms velocity
of the wiggly string on the scale of smallest wiggles found out to be,
\cite{stntwrk} $v_{st}\sim 0.6$. For this case, gravitational term can be 
neglected. In this case, density fluctuations for collisionless particles 
will be order of unity with opening angle, $\theta_w \simeq 8\pi G\mu_{w}$.
However, the velocity of the string obtained by taking average 
over a correlation length comes out to be very small ($v_{st}~\sim 0.15$) 
for which gravitational effect should be taken into account.
Formation of density structure of wake for collisionless particles have been
studied in ref. \cite{wiggle2} assuming the particles remain at a distance
far from the string and the particles will experience only averaged 
effect of the wiggles (in this case, relevant velocity is, $v_{st}~\sim 0.15$ ). 
In the string rest frame (string lying along z-axis), if the particles moves with
velocity $v_{st}$ in the (+ve) x direction, then following ref. \cite{wiggle2}, 
resulting density fluctuations can be written as, 

\begin{equation}
({\delta \rho \over \rho})_{axion} \simeq 1+ 4 G (\mu_w - T_w) ({{1 -v_{st}^2} 
\over v_{st}^2})({{x - X_0} \over X_0}) 
\end{equation}

Where, $X_0$ is of order of the inter-string separation. Note, setting the 
second term equal to zero, one recovers the result for density structure of the 
wake for straight string. If we put the value of wiggly string parameter in Eq.(15),
we see that order of magnitude of density fluctuations will not be affected 
compared to the case of straight string. We will see later that due to the trapping
mechanism, axion density can increase in the wake by many orders of
magnitude. We thus conclude that generation of density fluctuation
by axions (being collisionless) by above phenomenon is insignificant compared to
the case where axionic fluctuations happens through trapping mechanism inside the
wake produced by relativistic fluid. Therefore, we will ignore this effect in 
our subsequent discussion. 

  Now, we discuss the effect of wiggliness on the value of density fluctuations
(Eq.(13)) produced by moving string through the relativistic fluid. Study on 
formation of shock by moving wiggly string has been done 
in the work of ref. \cite{wiggle2} under some simplifying assumptions. In their 
study, authors have treated the fluid as a non-viscous, non-relativistic fluid. 
As has been argued in ref. \cite{wiggle2}, the assumptions of non-relativistic 
fluid can be good approximation for the string velocity $v_{st}\sim 0.2$ which
is relevant for the distance scale larger than the scale of the wiggles. 
Formation of wakes for wiggly strings has been discussed in ref. \cite{wiggle2}
after recombination. It is also mentioned in ref. \cite{wiggle2} that similar
wake should arise due to wiggly string even at earlier stages when string moves
through the relativistic plasma, even with $v_{st} = 0.15$ for the wiggly string.
If we follow the approach of ref. \cite{shk1,shk2,shk3}, then $v_{st} = 0.15$ is
subsonic and no shock can form. However, this does not represent the actual physical
situation. This is because at smaller distance scales (compared to the scale of the 
wiggles), the straight segments of the string move with $v_{st} \simeq 0.6$ 
(ref. \cite{stntwrk}). This is supersonic motion and shock will form. Due to random 
velocities of different segments of the string, the individual shocks will combine to
give some resultant shock and hence, wake for the total length of the (wiggly) string. 
Thus we will assume that even with the wiggles, shock and wake formation results.
For these wakes we will use Eq.(12) with $v_{st}= 0.6$ and mass per unit length $\mu$
for the straight string (as individual segments are straight). However, now the length
scale of the shock (wake) will be governed by the size of each straight segment.
The length of the wake (in the direction away from the string) will still be governed
by the average inter-string separation. That will not be affected by the presence of
the wiggles. (Recently, the evolution of wiggly string 
network has been studied \cite{vilenkin} on flat space-time. The results suggest that 
even in the presence of wiggles the string network obey scaling solution. Authors in 
ref.\cite{vilenkin} also mentioned that the average inter-string separation also scales with time. Though 
above simulation has been done on flat space-time, the author believes that such scaling
solution is expected to exhibit in expanding universe also. If this is the case, then 
on an average inter-string separation for long wiggly string will not be
altered from the straight string case.) However, the width of the wake (along the string 
direction) will now be truncated down to the average size of individual straight segment.
Note however that, this simply changes the geometry of a planar wake (for a completely
straight string) to a collection of strips with each strip corresponding to individual
straight segment of the wiggly string. Even the geometry of these strips may not be planar
due to rapidly changing velocity of string segments. For a correct treatment, one should 
properly account for the different velocity and orientations of these strips, however, 
for present we will ignore these complications. Following Eq.(12), the opening angle of 
the wake and density fluctuation for each straight segment of the wiggly string can now
be written as (using $v_{st} = 0.6$ and sound velocity  $v_s = 1/\sqrt{3}$),

\begin{equation}
\theta_w \simeq 70^0, \qquad  \delta \rho/\rho \simeq 4 \times 10^{-5} .
\end{equation}

As the individual shock will combine to give resultant shock, we expect that the overall
volume of the wakes will remain same (upto some factor) as for straight string case. 
Here, we should again emphasizing that the above value of density fluctuation and 
opening angle of the wake (i.e., Eq.(16)) is derived from Eq.(12), where it is assumed
that the fluid flow to be uniform. In reality, due to the wiggliness of the string the 
fluid will experience the rapidly changing directions of the wiggles which causes 
acceleration of the fluid in different direction. So, in proper treatment of shock 
formation by wiggly string, one should take the non-uniformity nature of the fluid 
in the analysis. However, due to lack of such complicated proper analysis, we will 
use (mainly) Eq. (12) and  Eq.(16) and  discuss the effect of such fluctuations on 
QCD phase transition.

\section{Effect of string wakes on quark-hadron transition and axion trapping
mechanism in our model}

 Here, we will first briefly discuss the dynamics of QCD phase transition in the
context of early universe. In the conventional scenario of first order phase
transition,  the transition happens through the nucleation of hadronic bubble
in the QGP background. These bubbles then grow, coalesce and finally convert
the whole QGP phase into hadronic phase. However, since the critical size of
the bubble beyond which it can grow is quite large at the temperature closed
to $T_c$ and also the nucleation rate too small at such temperatures, universe
has to supercool to a temperature $T_{sc}$ to begin the nucleation process.
The amount of supercooling $\Delta T_{sc}$ depends on the surface
tension $\sigma$ and the latent heat $L$. We take sample 
values of these parameters as obtained from the lattice results 
\cite{lattice}, $\sigma \simeq 0.015 T_c^{3}$ and $L \simeq 3 T_c^{4}$. 
As we discussed in ref. \cite{layek2}, for smaller value of surface
tension our results remain applicable as the supercooling would be smaller
(as has been argued in the literature \cite{gavai}, see also ref. \cite{venu}.) 
and cosmic string induced density fluctuations will have more prominent effect on 
the dynamics of quark-hadron transition.  Taking the values of these parameters one can 
estimate the amount of supercooling to be \cite{inhm,kjnt}

\begin{equation}
\Delta T_{sc} \equiv  1 - {T_{sc} \over T_c} \simeq 10^{-4} 
\end{equation}

for critical temperature $T_c = 150$ MeV.\\

As soon as the universe supercool down to temperature $T_{sc}$, hadronic
bubbles will start nucleating in the background of quark gluon plasma phases. 
The duration of this nucleation process is very small and lasts only for a 
temperature range of $\Delta T_n = (1 - {T_n \over T_c}) \simeq 10^{-6}$, 
for a time duration of order $\Delta t_n \simeq 10^{-5} t_H$ ($t_H$ is the 
Hubble time). The bubbles which have already been nucleated will keep  
expanding and it will release latent heat which will prevent further 
nucleation process. After the nucleation
processes stop, the universe enters into the so called slow combustion
phases \cite {wtn}. However, in our model \cite{layek1,layek2} this slow
combustion phase is very different from the standard scenario. Here the
phase transitions happens in the presence of density fluctuations (which
will transform into temperature fluctuations) which is produced as the string 
passes through the relativistic plasma.  In this scenario, as we discuss
below, the hadronization inside the wake will be delayed while outside
the wake the universe will enter into slow combustion phases. To understand
this slow combustion phase in our model we take density fluctuation from 
Eq.(13), which will transform into temperature fluctuation of magnitude  
$ \Delta T_{wake} \equiv \delta T/T \simeq 10^{-5}$.  Since, this temperature 
fluctuation is larger than $\Delta T_n $, there will be no nucleation inside
the wake while nucleation process will get completed outside the wake. 
Thus, the region outside the wake will enter into slow combustion phase, while 
overdensity region will till remain in QGP phase. For this, the overdensity
in the wake should not decrease in the time scale $\Delta t_n$. To calculate
the time scale $t_{shk}$ of the evolution of the overdensity we take the
typical average thickness of the wake as,

\begin{equation}
d_{shk} \simeq  sin\theta_w ~ d_H \simeq 3 km .
\end{equation}

Where, $d_H \simeq 10$ km is the horizon size at $T_c$ = 150 MeV.
(Note, as we have discussed earlier, the shock parameters of Eq.(16)
apply only for a straight segment of the string. However, the resulting wake may still
extend upto typical inter-string separation. Thus the overall volume covered by
the wakes of different segments of wiggly string could be of same order as that of a
wake from a straight string. We take this to be the case.) Now, typical time scale for 
the evolution of the overdensity will be governed by the sound speed $v_s$. We take 
$v_s \simeq 0.1$ suitable at temperature closed to $T_c$ (see ref. \cite{inhm,csmc}). 
Thus, we get the time scale as,

\begin{equation}
t_{shk} \simeq {d_{shk} \over v_s} \simeq 20 ~ sin\theta_w ~ t_H 
\end{equation}

which is large compared to the duration of 
the nucleation process $\Delta t_n \simeq 10^{-5} t_H$.  It is also much 
larger than $\Delta t_{trnsn} (\simeq 14 \mu sec)$ during which the quark 
phase is completely converted to the hadronic phase in the region outside the
wake. Thus, one can conclude that the region outside the wake enters the slow 
combustion phase before any significant bubble nucleation can take place in the 
wake region. Heat released by the bubbles outside the wakes will prevent bubble 
nucleation everywhere. It is then possible that the region outside the wakes
will be converted to the hadronic phase while inside the wakes QGP phase 
remains. Further completion of the phase transition will happen when the
interfaces, seperating the QGP phase inside the wakes from the outside
hadronic regions, move inward from the wake boundaries.
These moving, macroscopic, interfaces may trap most of the
baryon number in the entire region of the wake (and some neighborhood)
towards the inner region of the wake.  Finally the interfaces will merge,
completing the phase transition, and leading to a sheet of very large
baryon number density, extending across the horizon. Detailed profile of 
baryon inhomogeneities produced through these collapsing interface had been
estimated in our previous work \cite{layek2}.  
 
 Here, we want to discuss the mechanism of axions which will get trapped initially
inside the wake and subsequently leave after acquiring required momentum form the 
collapsing walls.  For that we will concentrate on a single wake like
overdensity region. Typical average volume of such region will be governed
by the total number of long strings in a given horizon. From numerical simulation
\cite{stntwrk}, the number of long strings is expected to be about 15. 
(As is mentioned earlier, simulation \cite{vilenkin} on wiggly string network suggests
that even in the presence of wiggles on long string, the network obey scaling solution.
Though above simulation has been done on flat space-time, the author
believes that such scaling solution is expected to exhibit in expanding universe also. 
If this is the case, then the number of long strings as we have taken for straight 
string case will remain applicable even for wiggly string also.) If the string wakes 
are reasonably parallel, then they will span most of the horizon volume, as the average
thickness of a wake  will be order of 1-2 km. In such a situation, the hadronic phase will
first appear in the regions between the wakes, which may cover a very small
fraction of the horizon volume initially.  If $f$ be the volume fraction occupied
by the QGP phase then the initial value of this fraction $f$  will be close to 1. 
$f$ will then slowly decrease to zero as the planar interfaces (formed by the 
coalescence of bubbles in the region in between the overdense wakes) move inward,
converting the QGP region inside the wake into the hadronic phase. Certainly, the 
actual situation will be more complicated than this, with string wakes
extending in random directions, and often even overlapping. In such a situation, 
even the initial value of $f$, when bubble coalescence (in the regions between 
the wakes) forming planar interfaces, may be smaller than 1. 
Note that, for wiggly string case, the wakes formed by straight segments of the string
will have different orientations depending on the direction of each straight segment. 
In this situation, distribution of string wakes will be even more complicated, hence in 
determining the exact initial value of $f$. However, for simplicity, 
we will take the initial value of $f$ to be almost 1 as is done in ref. \cite{layek2}
and focus on the region relevant for only one string, covering about 1/15 of the 
horizon volume. Also, as our main focus in this work is an order of magnitude 
estimate of iso-curvature fluctuations and since underline physical picture will
remain same even for wiggly string, hence from now onwards, we only consider the 
case of straight string and the results will be quoted only for straight string case.

 We now study the density evolution of axions which may get trapped initially
inside the wake. For that, let us first recall the effect of expansion of the 
universe on the dynamics of the phase transition. For that we mostly follow 
the work of Fuller \cite{fuller} (see also, \cite{layek1,layek2}) who have
studied the density evolution of baryons inside the shrinking QGP bubbles.  
If $R(t)$ is the scale factor of Robertson-Walker metric, then Einstein's 
equations give \cite{fuller},

\begin{equation}
{{\dot R(t)} \over R(t)} = \sqrt{{8\pi G \rho(t) \over 3}} ,
\end{equation}

\noindent where $\rho$ is the average energy density of the mixed phase 
and given by,

\begin{equation}
\rho= f\rho_q + (1-f) \rho_h
\end{equation}
The energy density, pressure ($\rho_q, p_q$) in the QGP phase and 
in the hadronic phase($\rho_h, p_h$) are given by,

\begin{eqnarray}
\rho_q =  g_q a T^4 + B, ~ p_q = {g_q \over 3} a T^4 - B \\
\rho_h =  g_h a T^4, ~ p_h = {g_h \over 3} a T^4 .
\end{eqnarray}

Here $g_q \simeq 51$ and $g_h \simeq 17$ are the degrees of freedom
relevant for the two phases respectively (taking two massless quark
flavors in the QGP phase, and counting other light particles) \cite{fuller}
and $a = {\pi ^2 \over 30}$. At the transition temperature we have 
$p_q = p_h$ which relates $T_c$ and the bag constant $B$ as,
$B = {1 \over 3} a T_c^4 (g_q - g_h)$. We define $x = g_q/g_h$ to be 
the ratio of degrees of freedom between the two phases. 
With this, Eq.(20) can be written as,

\begin{equation}
{\dot R(t) \over R(t)} = ({8\pi G B \over 3})^{1/2} 
[4 f + {3 \over {x -1}}]^{{1 \over 2}} .
\end{equation}

Now, conservation of the energy-momentum tensor gives,

\begin{equation}
R(t)^3 {dp(t) \over dt} = {d \over dt} \{ R(t)^3 [\rho(t) + p(t)] \} .
\end{equation}

During the transition, $T$ and $p$ are approximately constant. Therefore,
Eq.(25) can be rewritten as,

\begin{equation}
{{\dot R(t)} \over R(t)} = - {\dot f (x-1) \over 3 f(x-1) +3} .
\end{equation}

These two equations, Eq.(24) and Eq.(26) have to be solved simultaneously 
to determine the evolution of scale factor and the change of $f$, the volume 
fraction of QGP region as the transition proceeds.
The initial time ($t_0$) relevant for us is when the overdensity inside the 
wake has been formed. Taking 15 long string inside the horizon at that epoch,
we will consider the initial volume as,

\begin{equation}
V_0 \approx ({1 \over 15}) d_H ^3 ,
\end{equation}

\noindent where $d_H (= 2t_0)$ is the size of the horizon at time $t_0$.  
Note that we take the wake like overdense regions to be well formed at 
time $t_0$. Thus our representative volume as the transition proceeds is,
$V(t) = V_0 ({R(t) \over R_0})^3$, $R_0$ being the scale factor at $t_0$.
Taking center of the wake as the origin and considering motion of the interface along $z$ direction, we can write

\begin{equation}
f(t) V(t) = 2 A(t) z(t) .
\end{equation}
Where, $A(t)$ is the area of each sheet. Assuming the sheet extending
across the horizon, we get the area as a function of time as,
$A(t) \sim V(t)^{2/3}$. 
 Putting this value we get the evolution of the thickness $z(t)$ as,

\begin{equation}
z(t)= {f(t) \over 2} V_0 ^{({1 \over 3})} {R(t) \over R_0} .
\end{equation}

 Note that we are approximating the wake as bounded by two parallel sheets
separated by average thickness of the wedge.  Now let us determine the 
effect of interface motion on axion momentum distribution and subsequently
on the evolution of number density of axions inside the wake. Total number 
of axions which will remain initially inside the wake can be calculated by 
integrating Eq.(10). Thus, if $N(t_0)$ be the total number of axions trapped
initially inside the wake, then,

\begin{equation}
N(t_0)= V_0 {\int_{k_{min}}^{k_{max}} n_k dk}({R(\tilde t) \over R_0})^3.
\end{equation}

Where the last factor $({R(\tilde t) \over R_0})^3$ is due to  the decrease 
of axion density from time $\tilde t$ to $t_0$ causes by the expansion of the 
universe ($n_k$ as given in Eq.(10) gives the number density of axions
at time $\tilde t$). $k_{min}$ and $k_{max}$ are minimum and maximum comoving
momentum respectively. Since, the minimum comoving momentum which can be 
trapped initially inside the wake cannot be smaller than ${{2\pi R_0} 
\over z_0}$, ($z_0$ is the initial average thickness of the wake) we will 
take $k_{min}$ as  $\tilde {k}$ or $2\pi R_0/z_0$ whichever is larger. 
To determine that, we take the ratio,

\begin{equation}
{{\tilde k z_0} \over {2 \pi R_0}} = {R(\tilde t) \over \tilde t}
{z_0 \over {2 \pi R_0}}= ({1 \over {\tilde t t_0}})^{1/2}{z_0 \over {2 \pi}}
\end{equation}

Taking $\tilde t$ from Eq.(9), $t_0$ as $\sim 10^{-5} sec$ and the
initial thickness $z_0 \sim 1 $km or so, we get the ratio of $O(1)$.
So, we take the minimum comoving momentum $k_{min}$ as $\tilde k$.
The maximum comoving momentum $k_{max}$ of the axions which will remain
inside the wake can be set as discussed below. \\

The momentum of the axions will keep increasing with the collapsing 
interfaces \cite{hind}. Thus, if $k(t)$ be the comoving momentum of an 
axion whose initial momentum is $k_i$, then the momentum at any time t, 
when thickness of the wake becomes $z(t)$ can be obtained from the following
relation,

\begin{equation}
k(t)= k_i {z_0 \over z(t)} {R(t) \over R_0}
\end{equation}

The minimum physical momentum required for the axions to leave the
interface is equal to $\Delta m$ (see, Eq.(1)). Therefore, the axions which have
comoving momentum less than $\Delta m R_0$ will remain inside the
wake. Thus, we can put the upper limit of the integration in
Eq.(30) as, $k^*$ or $\Delta m R_0$ whichever is smaller. To 
determine that let us again calculate the ratio,

\begin{equation}
{k^*  \over { \Delta m R_0}} = {{R(t^*) \over t^*} \over { \Delta m R_0} }
= ({1 \over {t^* t_0}})^{1/2}{1 \over \Delta m }
\end{equation}

Putting the value of $\Delta m$ as obtained from Eq.(6) and Eq.(7) and the
value of $t_*$ from Eq.(4) the ratio can be determined in terms of the formation
scale of axionic strings as,

\begin{equation}
{k^*  \over { \Delta m R_0}} \simeq  7.9 ({t_0 \over sec})^{-1/2} 
({\eta_a \over {10^{12} GeV}})^3 \simeq 1.2 \times 10^3 ({\eta_a \over 
{10^{12} GeV}})^3
\end{equation}

Thus, the above ratio depends on the formation scale $\eta_a$ of the 
axionic string, which is constrained by the terrestrial and astrophysical
experiments as well as from the cosmological considerations. 
The most stringent lower bound has been obtained from the SN 1987A 
\cite{scale} as $\eta_a \ge 10^{10} GeV$ and the upper bound \cite{mass}
($\eta_a \le 10^{12} GeV$) comes from the consideration that the axions should 
not overclose the universe.  If we take $\eta_a < 10^{11} GeV$, then
the above ratio is always less than unity. 
In this case we can take the upper limit of integration in Eq.(30) as  $k^*$. 
Having set the limits of integration we can now calculate total number of 
axions which will remain initially inside the wake and given by, 

\begin{equation}
N(t_0)= V_0 F [{1 \over \tilde k} - {1 \over  k^*}] ({R(\tilde t) \over R_0})^3
\end{equation}

Where, $F= R(\tilde {t}) G \eta_a^{2} ln(\eta_a \tilde t) {\rho _c}$
$\simeq {3 \over {32 \pi}} \eta_a ^2 ln(\eta_a \tilde t) (\tilde t t_0)^{-1/2} 
{R_0 \over \tilde t}$.
Now as the interfaces  move towards each other momentum
of each axion will be modified according to Eq.(32). Let $n_k \prime dk
^ \prime$ be the modified spectrum due to trapping of axions within the
collapsing walls when thickness becomes $z(t)$. Unless the momentum of each 
axion becomes $\Delta m R(t)$ the axions will remain inside the wake. So upto
certain time $t_1$, total number of axions $N(t)$ (for $t_0<t<t_1$) will 
remain fixed inside the wake. Subsequently, number density will be increased
due to decrease of volume fraction of QGP region $f$ caused by interface
motion. So, total number of axions at certain time $t(<t_1)$ can be written
as,

\begin{equation}
N(t)= f V(t) {\int_{k ^\prime _{min}}^{k ^\prime _{max}} n_k^ 
\prime dk^ \prime} = f V(t) {\int_{k ^\prime _{min}}^{k ^\prime  _{max}} 
F^\prime {1 \over {k^\prime} ^2} dk^ \prime} \quad (t_0<t<t_1)
\end{equation}

Where, $F^\prime$ is new coefficient will be determined in terms of F. This
coefficient $F^\prime$ essentially will take care of the change of the 
coefficient F due to shifting of momentum of each axion as is obtained from
Eq.(32) and the change in QGP volume causes by interface motion. 
Determining this coefficient in terms F  will be particularly useful in 
determining the evolution of axion density when axions will start leaking out 
the overdensity regions.\\ 
$k^ \prime$ = $k {z_0 \over z(t)} {R(t) \over R_0}$ 
$\equiv$ $k c(t)$ (say) and $k^\prime_ {min}$,  $k^\prime _{max}$ have to be 
substituted by $\tilde k c(t)$ and $k^* c(t)$, respectively. Substituting all 
these quantities and integrating Eq.(36) we get the total number of axions 
at any intermediate time between $t_0$ to $t_1$. Equating this number to the 
initial total number of axions at $t_0$ as obtained from Eq.(35) one gets the  
coefficients $F ^\prime$ as follows,

\begin{equation}
F^ \prime = {{V_0 F c^2(t)} \over {f V(t)}} ({R(\tilde t) \over R_0})^3
\end{equation}

Putting back $F^\prime$ in Eq.(36) and divided by the QGP volume $f V(t)$
we get the number density of axions upto time $t_1$ and given by,

\begin{equation}
\rho(t)= {N(t) \over {f V(t)}}= {{V_0 F} \over {f V(t)}} ({R(\tilde t)
\over  R_0})^3  [{1 \over \tilde k} - {1 \over  k^*}]
= {3 \over {32 \pi}} {{V_0 \eta_a ^2} \over {f V(t)t_0^2}}ln(\eta_a \tilde t)
[{(\tilde t t_0)}^{1/2}- {(t_* t_0)}^{1/2}]
\end{equation}

Now, as we mentioned earlier that the density written above is
applicable upto time $t_1$ below which no axions will leave the QGP regions
due to insufficient momentum required to escape the region.
The thickness of the wake at that time $t_1$ can be calculated 
using the fact that when the momentum of axion exceeds the value 
$\Delta m R(t)$ they will leave the wake region, which is first happens
when $k^* c(t)$ becomes equal to $\Delta m R(t)$. Using this equality
we get the corresponding thickness at $t_1$ and determined by,

\begin{equation}
k^ \prime _{max} \equiv c(t_1)k^*= {z_0 \over z(t)} {R(t_1) \over R_0} 
{R(t^*) \over t^*} = \Delta m R(t_1)
\end{equation}

Or,

\begin{equation}
{z(t_1) \over z_0}= {1 \over \Delta m} (t^* t_0)^{-1/2} \simeq 7.8 
({t_0 \over sec})^{-1/2}({\eta_a \over {10^{12} GeV}})^3 \simeq
1.2 \times 10^{-3}
\end{equation}

Since, the initial thickness $z_0$ of the wake is $\sim 1 $km, and 
$t_0 \sim 10^{-5} sec$, thickness of the wake at time $t_1$ comes out 
to be  $z(t_1)$ $\sim 2 m$ for string formation scale $\eta_a \sim 10^{10}
GeV$. As soon as the thickness decreases down to $\sim 2 m$, the axions will
start leaving the QGP region and finally all the axions will leave the wake
and form a sheet like structure.  To determine the evolution of
density after  $t_1$, upper and lower limit of the integration of Eq.(36) 
should be replaced by  $\Delta m R(t)$ and $\tilde k c(t)$
respectively. Thus, the evolution of axion density within the wake after 
time $t_1$ is given by, 

\begin{eqnarray}
\rho (t)= {{V_0 F c(t)} \over {f V(t)}} ({R(\tilde t) \over R(t_0)})^3
[{1 \over {\tilde k c(t)}} - {1 \over {\Delta m R(t)}}] \\
= {3 \over {32 \pi}} {{V_0 \eta_a ^2} \over {f V(t)t_0^2}}ln(\eta_a \tilde t)
[{(\tilde t t_0)}^{1/2}- {z_0 \over {z(t) \Delta m}}], (t_1<t<t_f)
\end{eqnarray}

Where, Eq.(37) and Eq.(38) have been used. $t_f$ is the final time at which 
all the axions will leave the QGP region.  The corresponding thickness can be 
obtained from the following expression,

\begin{equation}
k^ \prime _{min} \equiv \tilde k c(t_f) = {z_0 \over z(t)} {R(t_f) \over R_0}
{R(\tilde t) \over \tilde t} = \Delta m R(t_f)
\end{equation}

So, the ratio of the final thickness when no axion will remain
inside to the initial thickness becomes,

\begin{equation}
{z(t_f) \over z_0} = {1 \over \Delta m} (\tilde t t_0)^{({-1/2})}
=8.31 \times 10^{-9} ({t_c \over sec})^{-1/3} ({t_0 \over sec})^{-1/2} ({\eta_a \over {10^{12} GeV}})^{5/6}
\end{equation}

Since, the time scale ($t_c$) at the onset of QCD phase transition is of
same order of magnitude as our initial time $t_0$, which is of the order
of $\sim 10^{-5} sec$  we get the ratio of final thickness to the 
initial thickness as $\sim 7.7 \times 10^{-7}$ for string formation scale
$\eta_a \sim 10^{10} GeV$. Therefore when the wake 
inside which axions were trapped reduced to a size of about $\sim 0.1 cm$, the axions
which were still remained inside the wake will all leave the wake
and form a sheet like region. The release of the axions from the wake
thus happens for the duration when wake thickness lies between 2 $m$ and 0.1
$cm$. Below the thickness of 0.1 $cm$, no axion will remain inside the
wakes. The axions thus left behind as the interfaces collapse will be 
concentrated in a sheet like regions. One can calculate the density profile
of these axions as follows.\\

Let, $\rho _{pr}(z)$ be the density of axions which is left behind at position 
z as the interfaces collapse.  Then we can write down the following expression 
which relates total number of axions at position $z$ to the density which are 
left behind as follows,

\begin{equation}
N(z) - N(z - dz) = A dz \rho_{pr} (z) ,
\end{equation}

\noindent where the time dependence of $z$ is given in Eq.(29). So We get,

\begin{equation}
 {dN \over dz} = A \rho_{pr} (z) .
\end{equation}

Thus we finally get the density profile of axions as a function of thickness
with collapsing interfaces as,

\begin{equation}
\rho_{pr} (z) = V_0^{({-2 \over 3})} ({R_0 \over R(t)})^2 ({dN \over dz}) .
\end{equation}

However, here we should mention that in deriving the above equation we
have considered that the axions will not disperse away immediately as it 
leaves the wake. Strictly speaking, the axions after leaving the wake will 
move with a velocity corresponding to the momentum they have in the hadronic 
phase, as they leave the wake. Since axions leave as soon as their momentum in 
the QGP phase equals the mass difference $\Delta m$ (Eq.(1)), their kinetic energy
may not be large in the hadronic phase. Hence they may not move very far from the 
wake (note, axions are very weakly interacting). In this case, the density 
will be somewhat decreased compare to the density as obtained from the 
Eq.(47). However the order of magnitude may not be changed much.  So for 
simplicity we will consider the case where the axions will not
be homogenized immediately after leaving the wake.
In the following section we will discuss the results as is obtained
from Eq.(36), Eq.(38) and Eq.(47).

\section {Result }

Eq.(24),(26) are solved numerically to get the evolution of scale factor
and volume fraction $f$ which is occupied by the QGP phase. The solution
thus obtained is fed into Eq.(38), and Eq. (42) to get the 
evolution of axion density within the wake as the transition from QGP
to hadron phase proceeds. Fig1. shows the evolution of axion density inside
the wake for two time zones. Time axis is given in the unit of $\mu sec$, while
density is in $fm^{-3}$. As is shown in the figure, most of the time during
which transition from QGP phase to hadronic phase ($\Delta t_{trans} \simeq
15 \mu sec$) happens, the axions will remain inside the wake. The reason is 
obviously due to ${1 \over k^2}$ dependence of axions density spectrum. 
The number of axions with minimum momentum will be more and they will leave 
the over density region last.
In Fig.1a, the density keeps increasing solely (for the time zone 
$(t_0<t<t_1)$) due to decrease in volume of QGP phase. While for next time
zone $ (t_1<t<t_f)$, there are two competitive processes. Decrease in
volume causes to increase the axion density, while leaking out of
axions will cause the density to decrease. Most of the time during which
transition from QGP phase to hadron phase proceeds, first effect will be 
dominated over the latter, hence there will be net 
increase in density which is evident from the Fig1b. Finally, almost all the
axion will acquire the momentum needed to escape from the QGP 
regions is achieved and density sharply decreases down towards
zero. The inset in Fig.1b shows the detailed profile of decreasing part of 
the axion density plot. Fig2 shows the evolution of density (in unit of 
$fm^{-3}$) with the thickness $z(t)$ (in unit of meter). Fig2a, shows the
plot upto thickness $z(t_1)$ until which density will keep on increasing 
due to the effect mentioned above. Fig2b, gives the evolution of density 
from thickness $z(t_1)$ to $z(t_f)$. For convenience, the inset in Fig2b. 
shows the expanded plot of the regions where 
density rapidly goes down towards zero. Fig.3 shows the plot of total number
of axions inside the wake as a function of thickness for the time zones,
$t_1 \le t \le t_f$. Here, volume of a wake is taken as the average
thickness of the wake multiplying by the area $A(t)$ (taking 15 long strings
per horizon\cite{stntwrk} and assuming the sheets extend across the horizon, 
area of each planar sheet is given as $A(t) \sim {({2 t})^2 \over 15}$) 
as discussed earlier. The inset in Fig.3 shows the expanded plot 
within very small distance interval to illustrate the decrease of the total number
of axions with the interface motions. Finally, the density profile of axions which
are left behind as the interface collapse which is obtained from Eq.(47) is 
shown in Fig4. Here, plot has been given for the relevant time interval 
$ (t_1<t<t_f)$. \\

 We can take average axion density of the order of $\sim 10^2$  
(in Fig.4), then we see that as the interfaces bordering the wake is
reduced to about $\sim 0.1$ cm, then almost all the axions will leave the wake 
(see Fig.3) and the density will be increased by a factor of $\sim 10^5$. 
Since, the energy of the axions required to escape from the wake regions comes
from the walls of the interfaces, these are kind of iso-curvature fluctuations.  
One can also calculate the total mass ($M_a$) of the axions which are 
concentrated within this planar sheet region (formed by left behind 
axions). Taking total number of 
axions $N_a$ of the order of $\simeq 10^{57}$ (See Fig.3), 
the mass of the axionic sheet can be obtained from Eq.(7) and given by,

\begin{equation}
M_q \simeq N_a~ m_{ah} \simeq 6.04 \times 10^{-6} eV ({{10^{12} GeV} 
\over \eta_a }) N_a \simeq 10^{19} ({{10^{12} GeV} \over \eta_a }) gm
\end{equation}

Taking $\eta_a \simeq 10^{10} GeV$, the upper bound on mass of the 
axionic sheet will be order of $10^{21}$ gm ($\sim 10^{-12}M_{\odot}$ at QCD
epoch). Since, the fluctuations produced by axion trapping are of iso-curvature
kind, they grow little \cite{hind,rees} before radiation-matter equality era, 
$t_{eq}$. After $t_{eq}$, they grow in proportional to the scale factor.
Hogan and Rees \cite{rees} have studied the evolution of iso-curvature 
fluctuations which were produced at the era of QCD phase transition. 
They have shown that iso-curvature perturbations produced by axions at 
QCD epoch can lead to formation of axionic 'minicluster'. 
The amplitude of the iso-curvature fluctuations produced in our model are 
very large and axions are concentrated within a very narrow sheet like regions
of thickness of order  0.1 cm. This sheet will extend to a distance scale
of order $\sqrt A(t)$. Where, $A(t) \simeq {(2t)^2 \over 15}$ is the typical area
of the wake as discussed above. This is about 2 km at QCD scale, which corresponds
to comoving distance scale of order $10^{-7}$ Mpc today. One can study the evolution
of such sheet like overdense region as it enters into the radiation-matter equality
era. One can also study these overdensity at larger distance scales resulting
from the large scale distribution of strings and their wakes. If these fluctuations 
survive until late stages, it will be interesting to study the effects of sheet
like axionic clusters, especially whether they can have any effects on small scale 
CMBR anisotropies. 

\section{Conclusion}

In this paper, we have studied the iso-curvature fluctuations in
axion density at QCD phase transition epoch due to the presence of density 
fluctuations produced by moving cosmic strings. We have considered the axions 
which are produced from the radiation of the axionic strings which are formed
at some scale $\eta_a$ due to breaking $U(1)_{PQ}$ symmetry. If the mass
of the axions is relatively higher in the hadron phase compared to QGP phase
then the axions may get trapped initially inside the wake-like overdensity 
regions. As the transition from hadronic phase to QGP phase proceeds with 
the motion of interfaces, these axions will acquire momentum due to
collapsing interfaces and subsequently leave the wake. The axions thus 
left behind as the interfaces collapse may produce iso-curvature 
fluctuations. We have estimated the detailed profile of the fluctuations. We 
have shown that, in our model the iso-curvature fluctuations in the axion 
density will be of order $({\delta \rho \over \rho})_{axion} \sim 10^5$ and 
they will be concentrated within a planar sheet like region of thickness few cm. 
This sheet can extend upto a distance scale of order 2 km at QCD
scale. This length scale corresponds to comoving length scale of order 
$10^{-7}$ Mpc today. It will be interesting to  study the implications
of such large fluctuations in the axion density especially on small scale CMBR 
fluctuations. Here, we should mention that, though we have discussed the density
structure of wake for straight strings as well as for perturbed strings, results 
are quoted for straight string wakes only. For wiggly strings, due to the lack of 
proper relativisitic analysis of shock formation, we have followed the analysis of 
ref. \cite{shk3} and quote the resulting density fluctuations and opening angle
of the wake (Eq.(16)) for wiggly string. The formation of shock in ref. 
\cite{shk3} has been studied for straight string moving through relativistic
fluid. In their analysis, the fluid flow has been taken to be uniform. This 
assumption may not hold for wiggly string. This is because, at the presence
of wiggles, the fluid will experience the rapidly changing directions of the 
wiggles which causes acceleration of the fluid in different direction. So, in 
proper treatment of shock formation by wiggly string, one should take the 
non-uniformity nature of the fluid in the analysis. 
We have also discussed the case where axion (being collisionless) directly 
(other than trapping mechanism) can give rise to density fluctuations. It turns
out that, the value of these fluctuations are very small compared to the 
fluctuations produced from trapping of axions by cosmic string wakes.

\vskip .2in
\centerline {\bf ACKNOWLEDGEMENTS}
\vskip .1in

  I am very thankful to Ajit M. Srivastava for many useful comments 
and discussions. I am also thankful to Soma Sanyal, Rajarshi Ray and 
Ananta Prasad Mishra for many useful suggestions and comments. 


\newpage
\begin{figure}[h]
\begin{center}
\leavevmode
\epsfysize=11truecm \vbox{\epsfbox{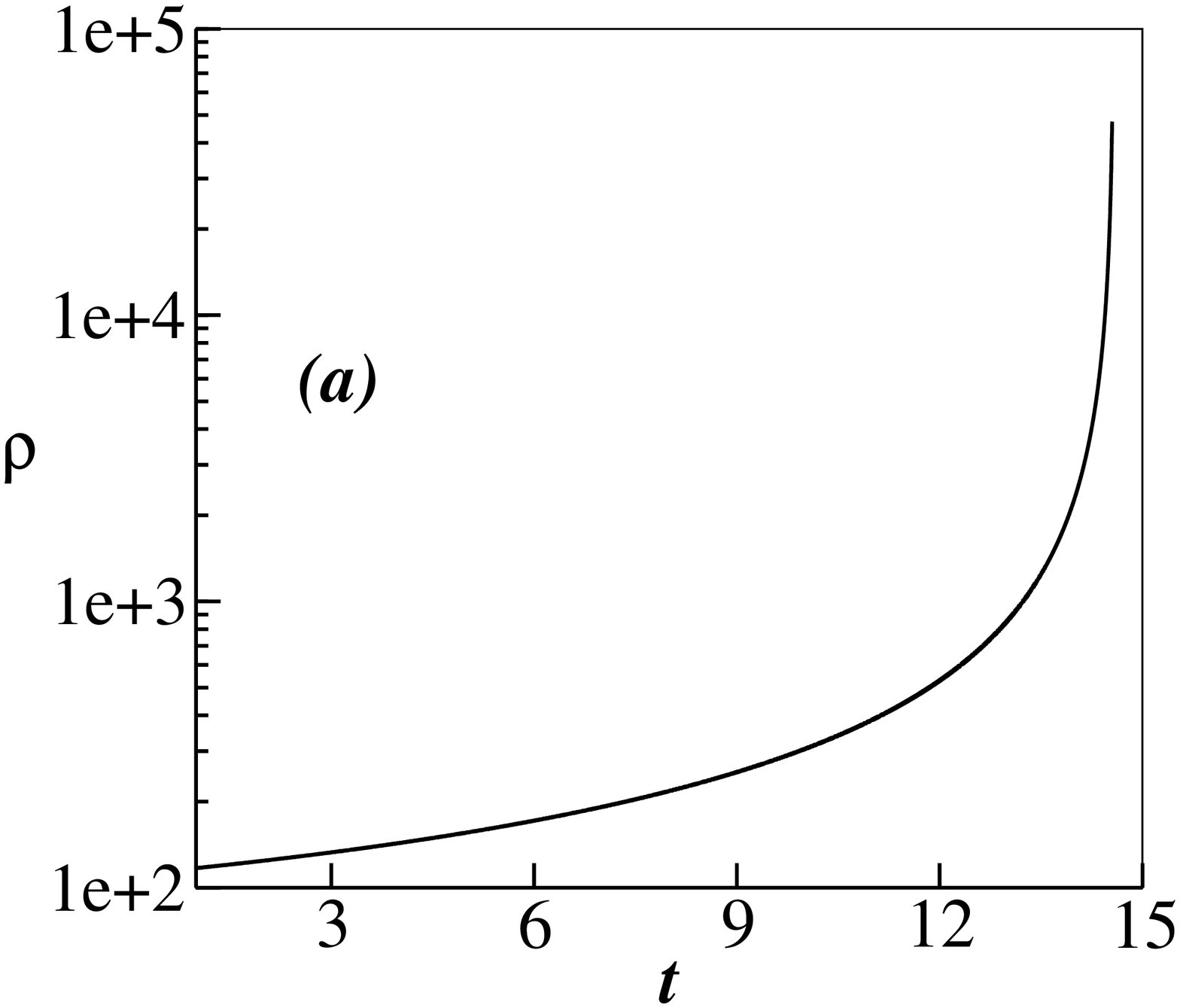}}
\epsfysize=10truecm \vbox{\epsfbox{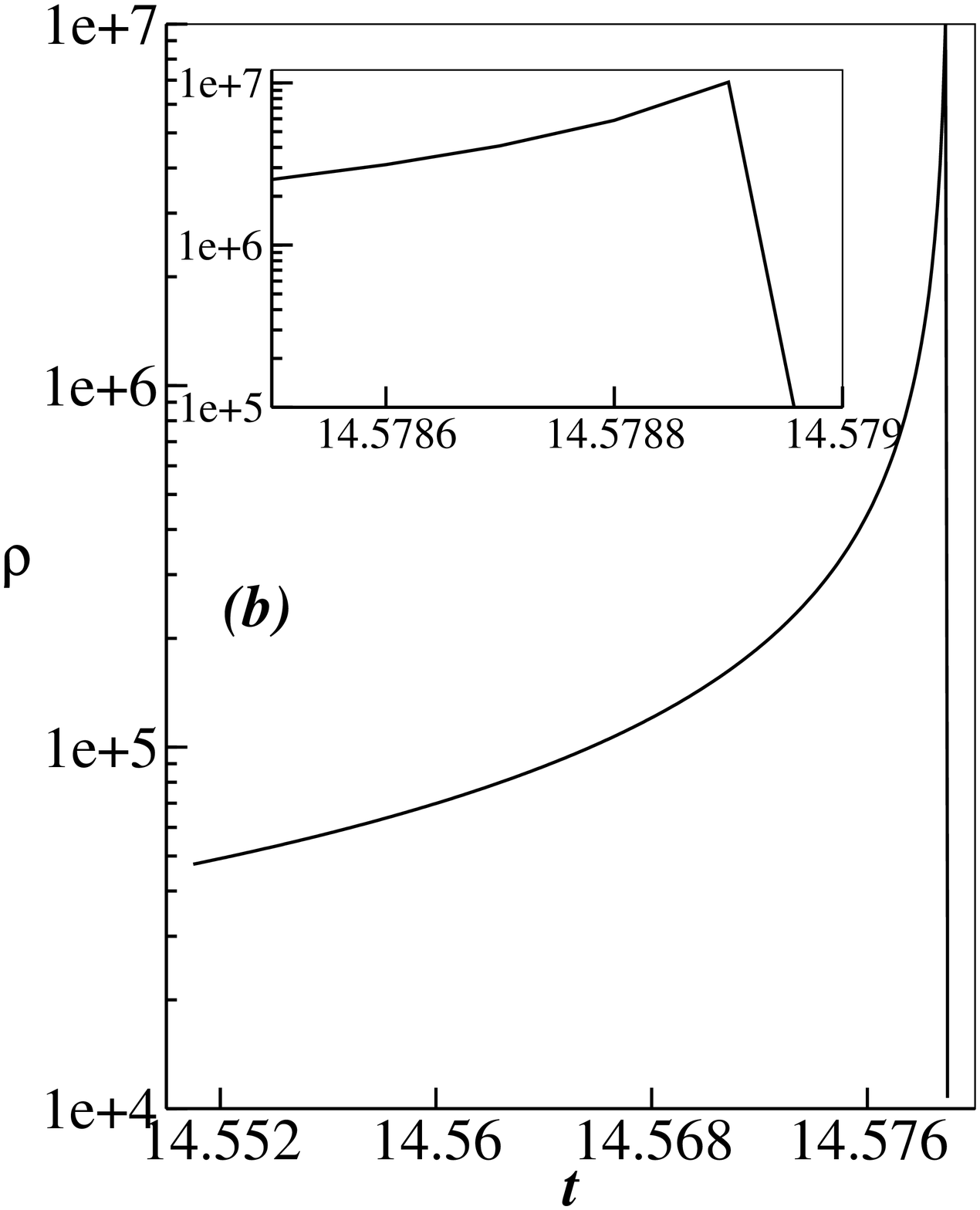}}
\end{center}
\caption {These figures show plot of evolution of axion density inside the
wake as a function of time. The density $\rho$ is given in unit of $fm^{-3}$
and time $t$ is in $\mu sec$. Figures (a) and (b) show the density plot for two
time zones $t_0 \le t \le t_1$ and $t_1 \le t \le t_f$ respectively.}
\label{Fig.1}
\end{figure}

\newpage
\begin{figure}[h]
\begin{center}
\leavevmode
\epsfysize=10truecm \vbox{\epsfbox{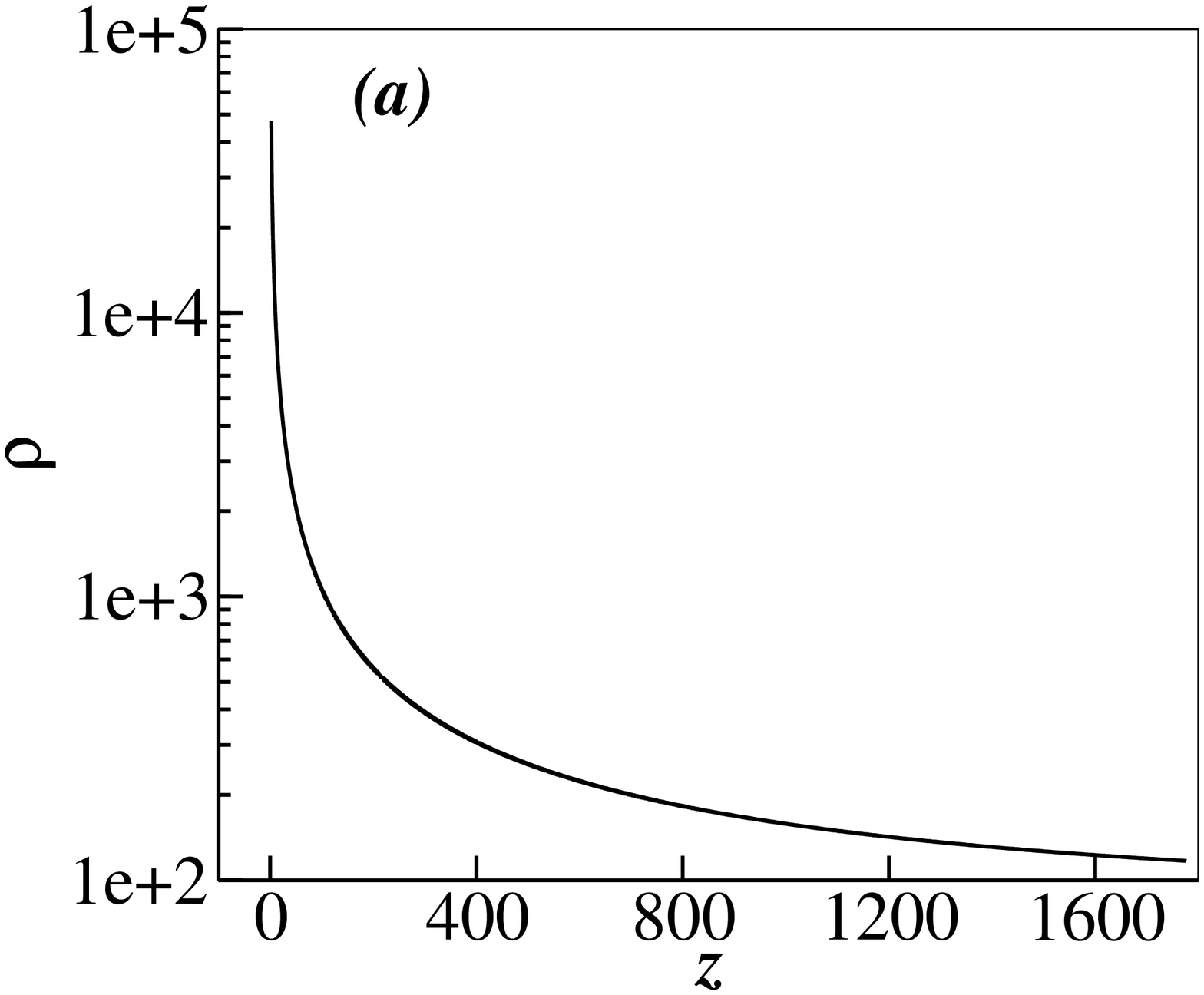}}
\epsfysize=10truecm \vbox{\epsfbox{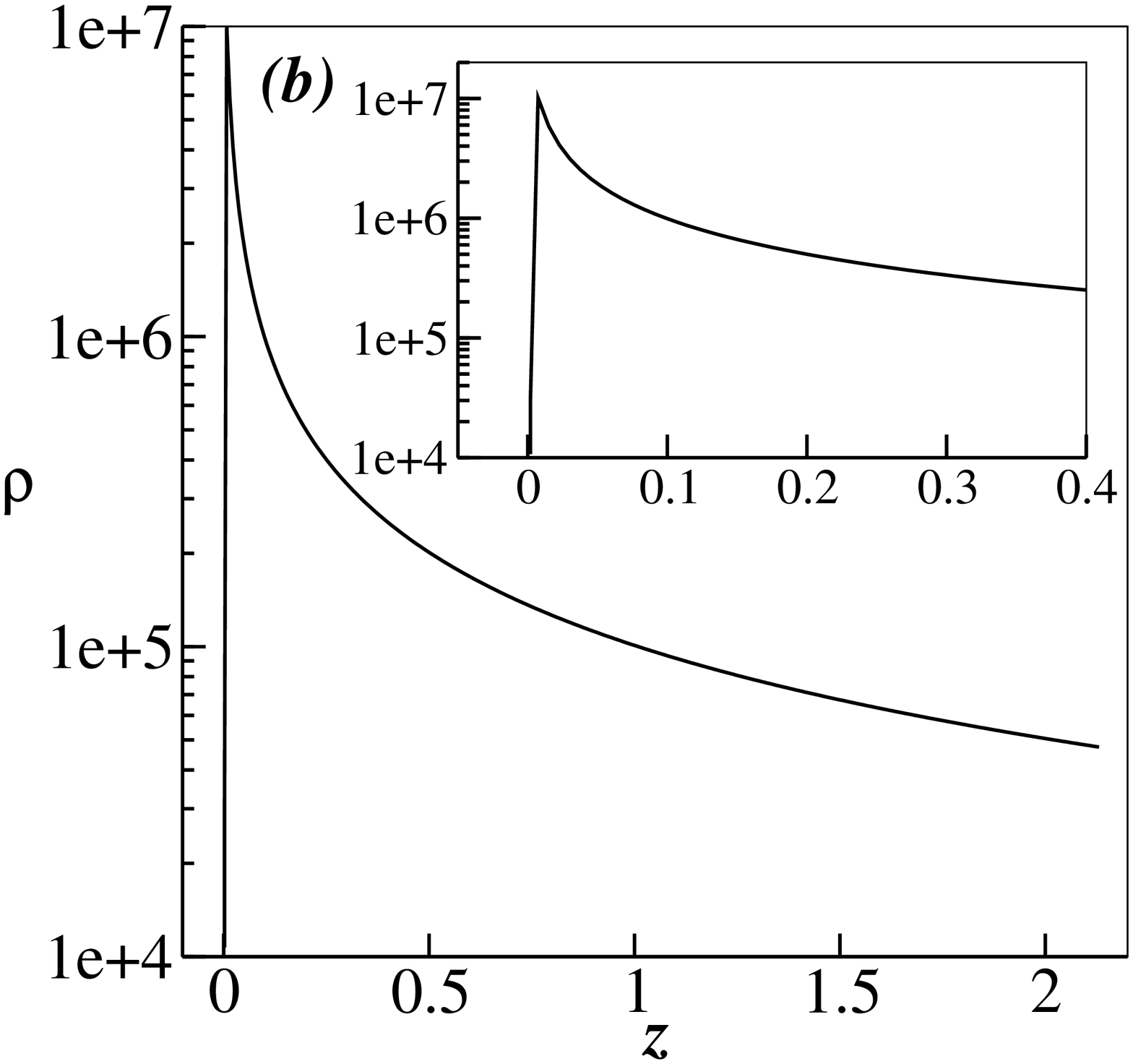}}
\end{center}
\caption {These figures show plot of evolution of axion density $\rho$ 
(in unit of $fm^{-3}$) inside the wake as a function of thickness $z$ 
(in $meter$) of the wake. Fig.(a) shows the plot upto thickness $z(t_1)$
and Fig.(b) is from $z(t_1)$ upto $z(t_f)$.}
\label{Fig.2}
\end{figure}
\newpage
\vskip -0.25in
\begin{figure}[h]
\begin{center}
\leavevmode
\epsfysize=10truecm \vbox{\epsfbox{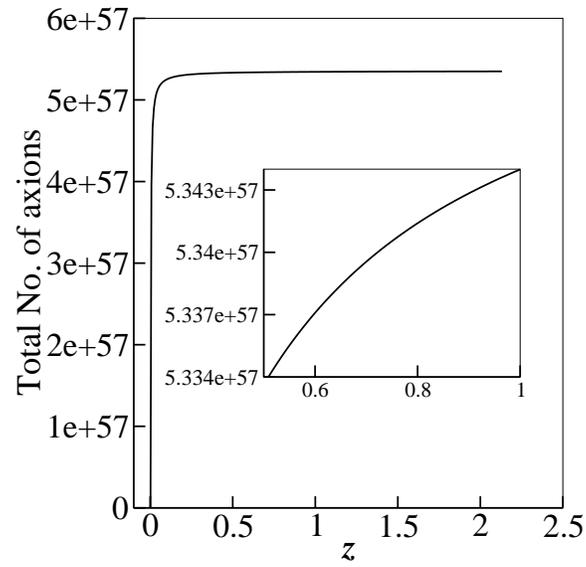}}
\end{center}
\vskip -5in
\caption {This figure shows the decrease of total number of axions during 
the time interval $t_1<t<t_f$ as a function of thickness $z$ (in $meter$). 
Small part of the apparently flat portion of the plot is expanded in the 
inset.}
\label{Fig.3}
\end{figure}
\newpage
\vskip -0.25in
\begin{figure}[h]
\begin{center}
\leavevmode
\epsfysize=11truecm \vbox{\epsfbox{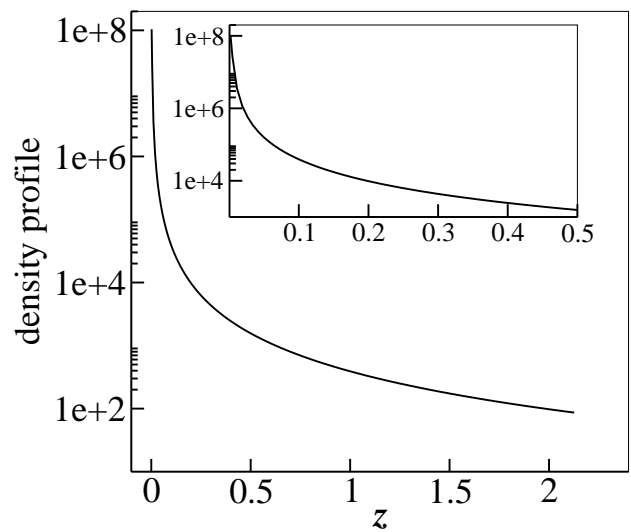}}
\end{center}
\vskip -5in
\caption {This figure shows the density profile of the axions which are
left behind as the interfaces collapse. Density is given in unit of $fm^{-3}$
and thickness $z$ in $meter$.}
\label{Fig.4}
\end{figure}

\end{document}